# POWER GENERATION ON A SOLAR PHOTOVOLTAIC MODULE INTEGRATED LIGHTER-THAN-AIR PLATFORM AT A LOW ALTITUDE

[1]Kuntal Ghosh, [2]Anirban Guha, [3]Siddharth P. Duttagupta

I.I.T Bombay

## ABSTRACT

Use of lighter than air platforms (aerostats and airships) for reconnaissance and surveillance over long periods can be facilitated by generation of power on-board through solar photovoltaic arrays. Attempts to integrate solar photovoltaic modules on such contoured surfaces leads to multiple challenges ranging from the choice of solar modules, determination of best method of their integration with the lighter than air platform, and design of the array layout in order to optimize the loss of power due to non-uniform illumination. This paper describes the method of designing such a system and suggests strategies for overcoming these challenges. The issue of non-uniform illumination has been tackled by maximum power point tracking using the scanning window technique for maximizing power generation and an algorithm of distributed maximum power point tracking has been suggested for further improvement. The procedure described in this paper can be used for obtaining the optimum power generation capability of a solar photovoltaic array mounted on a lighter than air platform of a given volume and payload carrying capability.

## INTRODUCTION

There has been a renewed interest in aerostats and airships in recent years due to their low cost of deployment for extended periods compared to fixed wing aircrafts and quad-rotor drones. Their use is being explored for reconnaissance and surveillance at coastal and international border areas, communication link setup in remote areas, exploration of rare earth minerals and water for geological survey and geographical information system setup for geographical mapping of forest, rural, urban and semi urban areas. Steady and uninterrupted electrical power supply is required for these applications. This requires power to be generated in situ, i.e. on the airship or aerostat, preferably through a solar photovoltaic (SPV) module.

The concept of power generation at a low altitude has first been explored by Glaser et al. [2. Glaser, 74] in 1970. He investigated the feasibility of harvesting solar energy using large satellites and transmitting it to the ground via microwave radiation which could be converted to electrical energy for the further use [2. Glaser, 74]. This idea has been explored by Mankins et al. for low earth orbit (LEO) satellites and geo-stationary satellites (GEO) [3. Mankins, 97]. Complications arising from Doppler effect between the satellites and the ground station while harvesting and transmitting the space based solar power (SBSP) has also been investigated [4. Goh, 12]. However technical issues regarding losses in the energy conversion and transmission, safety concerns (while microwave beam linking between satellite and ground) and implementation cost have resulted in scarce implementation of this idea. Aglietti et al.[5. Aglietti, PIP, 08] overcame some of these problems by proposing a high altitude aerostatic platform (HAAP) at 6 km altitude to harness solar power [6. Aglietti, JAE, 08]. Since this HAAP generates power above the clouds and transmits it to the ground through a cable, it gets the advantage of higher solar irradiance compared to ground based PV systems. This has a significant advantage in high latitude countries where peak sun hours are only between 2.5 – 4.85 hours [5. Aglietti, PIP, 08, 7. Aglietti, 10]. Further research in this area explored power transmission losses, control strategy for positioning the aerostat at a certain location and orientation along with its economic viability [5. Aglietti, PIP, 08, 7. Aglietti, 10, 8. Aglietti, *WSEAS*, 08, 9. Aglietti, 09]. Venusian atmosphere exploration [10.Hall, 08] and stratospheric aerosol geoengineering [11.Davidson, 74] has also been investigated as applications of this concept. Aerostat mounted PV generated solar energy powered reverse osmosis for cleaning water and emergency power supply in remote areas [12. Garg, 15] has also been reported. Spherical aerostats have been studied in all these cases in spite of their low aerodynamic stability.

High altitude PV power generation has also been investigated in the context of multibody advanced airship for transport (MAAT). This would draw operational energy for propulsion and control through integrated solar power – fuel cell hybrid power system [13. Khoshnoud, 13]. This concept has been further elaborated by Chang [14. Chang, 13] to explore the power generation characteristics on SPV modules integrated on aeroplane wings. Power generation at high altitudes using heavier than air structures has been studied by Yashwanth et al[15. Yashwanth, 12]. They have proposed harvesting of wind energy at high altitudes using wind turbines. These flying electric generators (FEG) collect wind power from heights ranging from 183 m to 10 km by exploiting powerful jet stream of currents [15. Yashwanth, 12, 16. Ferguson, 10]. But harnessing this energy is completely dependent on winds [17. Roberts, 07]. A problem arises when this does not get enough wind force to support its structure at a high altitude and starts acting like a heavier than air system. Instead of supplying power to the ground, it starts drawing power from the ground. Magenn air rotor system overcomes this problem by installing the machines on a helium filled buoyant structure [16. Ferguson, 10]. This system has also failed to gain popularity, perhaps due to the complexity of the energy conversion process. Use of solar PV modules seems to have an advantage over such systems.

The use of spherical aerostats, as investigated in the systems described above, might be acceptable in the stratospheric range. Low air drag at that rarified atmosphere may allow this to be a feasible design. However, presence of high wind at lower atmospheric ranges, need the effect of high air drag and its associated problems to be addressed. The displacement of the aerostat from its desired position (blow by) [18. Kale, 05, 19. Ram, 10] needs to be studied and minimized. An aerodynamically efficient contoured G. N. V. Rao shaped aerostat would be a better option for low altitude operations. Hence, an aerostat of this shape has been designed as an SPV integrated lighter-than-air platform and has been studied in this work.

This however requires the solar PV array to be mounted on a contoured platform. The study of such a system is the primary focus of this work. Previous research in this area includes a feasibility study by Sharma et al [28. Sharma, 10] of power generation and integrating an amorphous-Si solar module on a curved surface [20. Gajra, 14]. They have explored the effect of surface contour angle on power generation of an amorphous-Si SPV module. But this phenomenon is yet to be explored in the context of the contour of a lighter than air platform

(LTAP) to estimate its maximum power generation ability. Non-uniform illumination (NUI) is inevitable which effects the power generation on the SPV integrated LTAP. A bypass diode should be integrated with each and every module in parallel to mitigate mismatches due to NUI and partial shading. This arrangement increases the power output due to the increase of the module current in a fixed state. Meanwhile it causes a shift in operating voltage to a lower value by $\Delta V$ [21. Patnaik, 12]. Hence an alternative scheme is proposed by Nimni et al. [22. Nimni, 10] where a distributed DC-DC converter is used for power optimization due to mismatches in the SPV array. The effect of NUI on SPV array power generation has been solved by different reconfiguration strategies which have been reported by a number of researchers [23. Auttawaitkul, 98, 24. Nguyen, 08, 25. Cheng, 10, 26. Velasco-Quesada, 09, 27. Salameh, 90]. These reconfiguration strategies have been implemented in a fixed state SPV array on a planar region, not on a contoured surface like an LTAP. Later this concept has been further explored on the contoured plane using amorphous-Si flexible PV modules [28. Sharma, 10, 29. Sharma, IA, 14, 30. Sharma, 09]. These articles describe the effect of contour on the power generation and explain the methodology for determining maximum power generation using scanning window technique (SWT). Different distributed maximum power point tracking (DMPPT) architectures are explored under mismatching conditions due to partial shading which enable incident NUI on the SPV array. Those techniques include voltage equalization by using SEPIC configuration [31. Pragallapati, 13], current equalization by connecting DC-DC converters in parallel with each PV modules in the array [32. Sharma, 13, 33. Sharma, PE, 14] and full power processing architecture, minimum power processing architectures [34. Shmilovitz, 12]. These DMPPT techniques can also be implemented in the sub module level by integrating sub-module integrated converters which eliminates potential mismatches due to ageing, NUI and partial shading [35. Olalla, 14]. Moreover performance of these distributed DC-DC converters can be estimated theoretically against the mismatches [36. Grasso, 15]. However, the effect NUI in the context of the contoured surface of an LTAP (aerostat) and the strategy for maximizing power generation in such a system has not been studied till date. This is the focus of the current study.

This paper explores the feasibility of optimized electric power generation on an LTAP. First the generic methodology to construct such an electric power generating platform is described which includes sizing of the platform. Then envelope contour function (ECF) of that platform is briefly explained and incident non-uniform illumination at different ECF is demonstrated. This leads to solar power generation and its maximum power tracking. With the findings in maximum power on the LTAP, distributed maximum power point (DMPPT) technique in the sub module level is discussed and recommended for the optimized power generation.

**EXPERIMENTAL SETUP**

The present research aims to design and develop a mobile airborne platform which is capable of power generation and subsequently develop a DC microgrid for remote area power supply. This requires the design of an aerodynamically stable platform and its appropriate sizing to support the power system and associated payloads at a certain altitude. This needs consideration of additional safety factor in the design stage. The proposed system is demonstrated in figure 1.

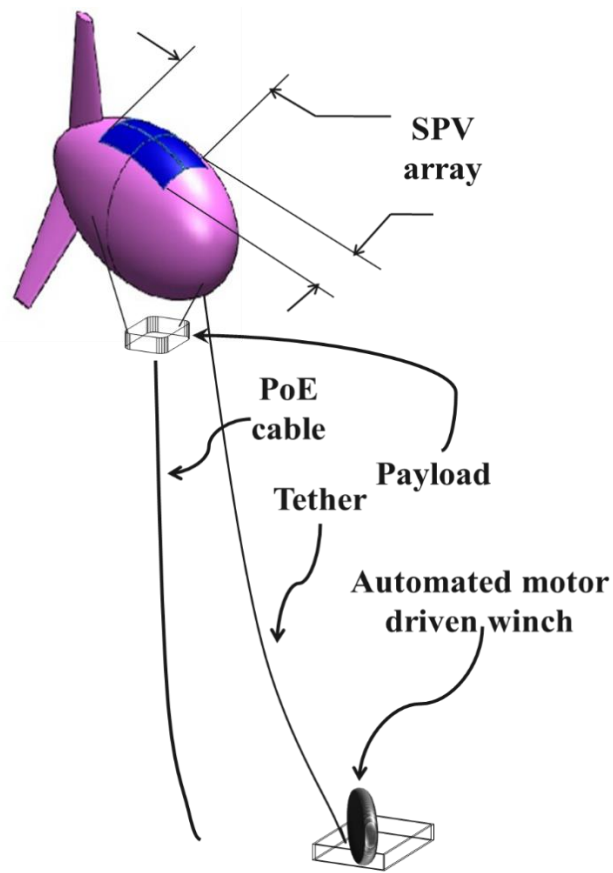

**Fig. 1** System diagram

This entire process requires series of methods which are depicted below:

- <u>Selection of the power source:</u> There are different kinds of commercially available flexible solar cell technologies. The primary selection criteria are solar PV technology (reliability, altitude dependent performance), Watts/kg (maximum power/weight) and Watts/m$^2$ (maximum efficiency). Table 1 summarizes different solar PV technologies and their parameter dependencies.

**Table 1 :** Comparison of different SPV technologies

| Manufacturer | SPV Technology | Peak power ($W_p$) | Area ($m^2$) | Weight (kg) | Current, Voltage @ $W_p$ | Watt-p /kg | Watt-p /$m^2$ |
|---|---|---|---|---|---|---|---|
| Global Solar *PowerFlex* [38. Global Solar, 16] | Copper Indium Gallium (di) Sulfide/ Selenide (CIGS) | 300 | 2.843 | 9.3 | 5.5 A 54.3 V | **32.25** | **105.52** |
| High-Flex Solar HF315 [39. HF315, 16] | Mono-crystalline Silicon | 315 | 2 | 3.6 | 8.62 A 36.5 V | **87.5** | **157.5** |
| High-Flex Solar HF40 [40. HF40, 16] | Mono-crystalline Silicon | 40 | 0.315 | 0.7 | 2.48 A 16.16 V | **57.14** | **126.9** |
| Power Film Solar FM16-7200 [41.PowerFilm, 16] | Amorphous Silicon | 120 | 1.26 | 2.95 | 7.2 A 15.4V | **40.68** | **39.47** |

Comparison of the Watt-p/kg and Watt-p/m² values leads to the conclusion that High Flex solar HF – 40 SPV modules are most suitable for this system.

- <u>Sizing of the structural unit:</u> Sizing of the structural unit is performed based on the required payload. An aerodynamically stable G.N.V.Rao shaped envelope made of synthetic fabric has been filled by helium gas which creates an upward buoyancy force which lifts it up to its pressure height. This structure has a certain payload capacity which can be determined by the following mathematical equations.

$$B = V \rho_a \quad \text{-------------- (1)}$$

Where B is the upward buoyancy force acting on the LTAP

$\rho_a$ is the mean density of the local atmosphere surrounding the aerostat

Then the net upward lift force (L) = B – W  ---------------- (2)

Where W = Weight of the inflated envelope = $V \rho_g + W_0$  ---------------- (3)

$\rho_g$ = Density of LTA (Lighter than air, $\rho_a > \rho_g$) gas

$W_0$ = Weight of the envelope + weight of hooks and patches + weight of nose battens + weight of gas filling port+ weight of ballonet

So, Equation (5) can be rewritten as $L_d = V.(\rho_a - \rho_g) - W_0$  ---------------- (4)

$$L_d = L_g - W_0 \quad \text{---------------- (4.a)}$$

Here  $L_d$ = Disposable lift of the LTAP envelope

$L_g$ = Gross lift = $V. (\rho_a - \rho_g)$

Hence maximum payload capacity of an inflated LTAP = $L_d$

It is estimated that applications ranging from …. to ……. would require about 160 $W_p$. Thus, a prototype of 160 $W_p$ power system was developed which has the effective payload of 4 x 0.7 = 2.8 kg. It would require an LTAP of minimum 24 m³ volume to support this effective payload. Details of this structure are as follows:

Maximum diameter   : 2.4 m

Maxium length         : 7.32 m
Volume                : 24 m$^3$
Surface area          : 43 m$^2$

- Integration of SPV modules to the structure:  These modules are integrated on the LTAP using Velcro straps. Silicon glues are found to be the best proven technology to integrate male straps with the SPV modules and female straps at the specified location on the LTAP. This method of connection is
    I. Flexible in nature
    II. Produces no thermal effect
    III. Provides excellent resistance to oil, chemicals and moisture
    IV. Waterproof

- Envelope contour function (E.C.F): SPV array produces a variable tilt while mounting the flexible SPV modules on the LTAP. These tilts are a function of the envelope size which is termed as envelope contour function (ECF). Identical ECFs are generated on both sides of the LTAP envelope since it is axisymmetric.

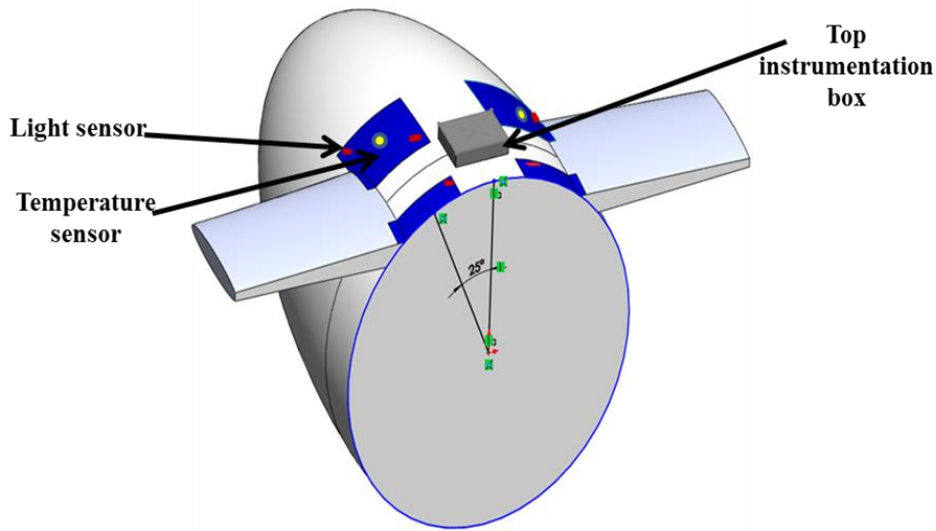

**Fig. 2**  Envelope contour function of the prototype

Average $25^0$ ECF is created at the SPV layout on the both sides of the prototype envelope as seen in Fig. 2.

- <u>Maximum power generation employing an automated variable load and data acquisition system:</u> A setup has been established for data acquisition from the sensors as well as for maximum power point tracking based on certain conditions. The setup consists of an instrumentation box which has been kept on top of the LTAP as shown in Fig. 2. The acquired data has been transmitted to the bottom data acquisition box on ground through a RS-232 cable. Details of the process is given below:

The top instrumentation box is used to switch different combinations between the solar PV modules. There are 3 combinations - series, parallel and series-parallel. One temperature sensor (RTD – PT100) and a light sensor (LDR) are attached to each solar PV module to measure their temperature and incident irradiance respectively. These sensor output signals are sent to the bottom data acquisition box. A schematic diagram is shown in figure 3(a) and the components are shown in figure 3(b).

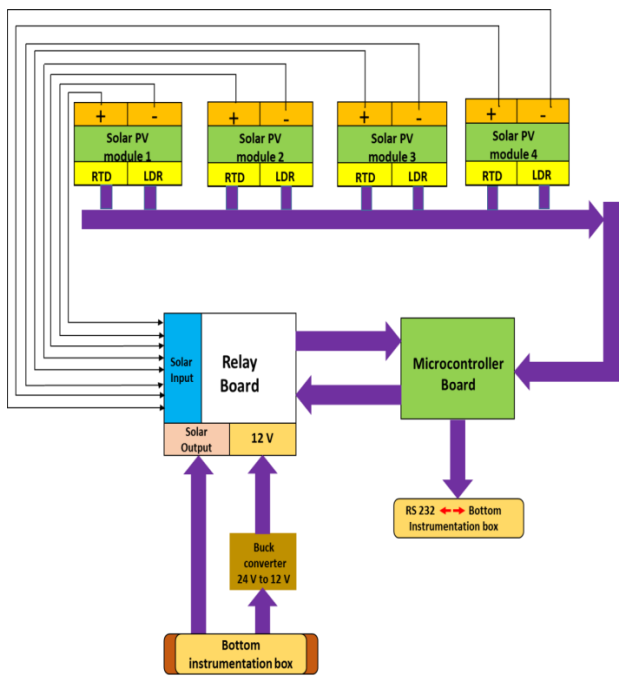
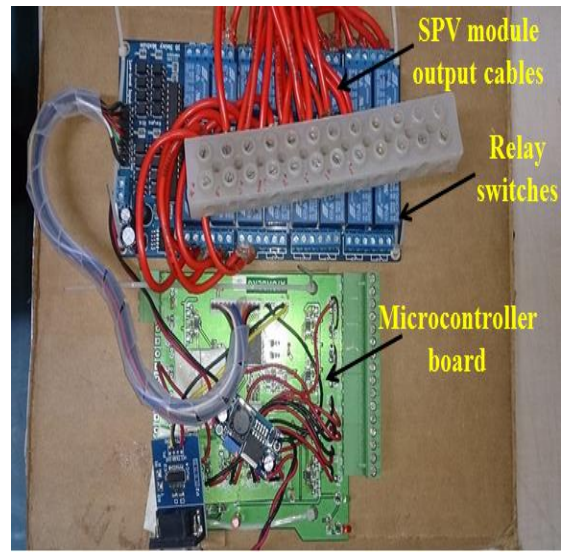

**Fig.3 (a)** Block diagram of top instrumentation box        **Fig.3 (b)** Top instrumentation box

Each solar modules generated voltage (solar input) given to the relay board. Power drawn from the bottom microcontroller board (24V) is supplied to the buck converter. It provides 12V power supply for the relay board. The top microcontroller board gets 5V supply from relay board buck converter. In addition it also receives control signal from the other microcontroller which is situated in the bottom instrumentation box and controls the relay board accordingly. It also measures the temperature and the incident solar irradiance to each sensor integrated with every SPV modules. It sends the real time data to bottom data acquisition box. 24 V power supply is used to power the bottom microcontroller, stepper motor driver and microcontroller in the top instrumentation box. Microcontroller recieves instructions from application software through the usb cable. Bottom microcontroller receives temperature and illuminance data from top instrumentation unit through the RS-232 cable and forwards it for the further application and processing. A schematic diagram is shown in figure 4(a) and the components are shown in figure 4(b).

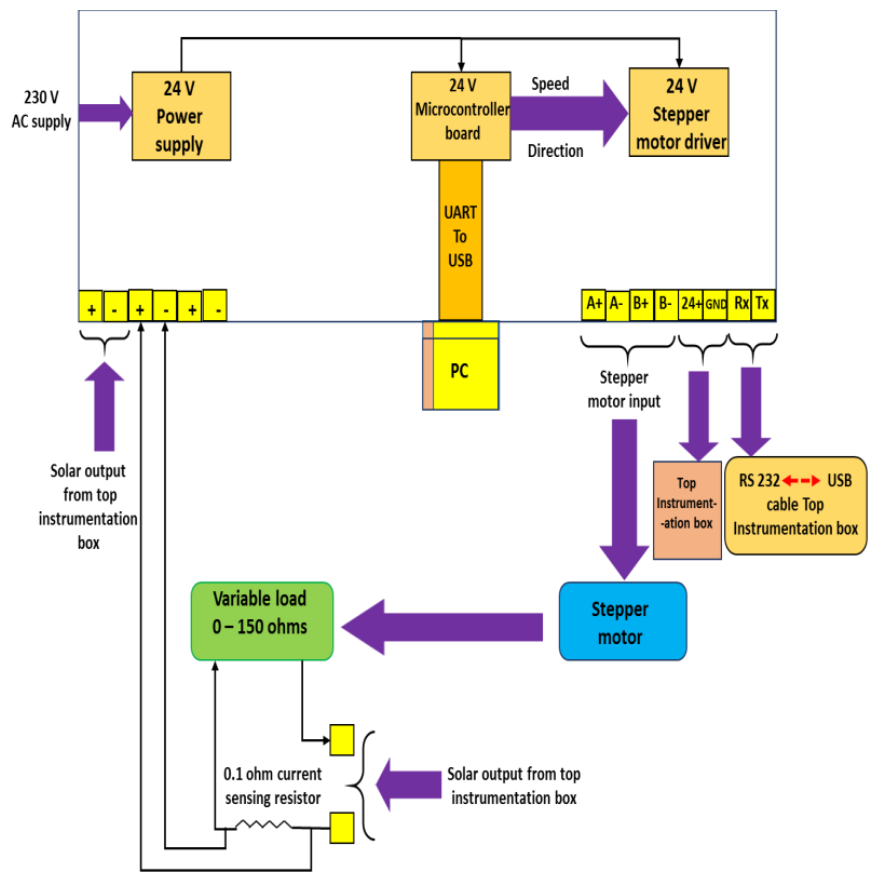

**Fig.4(a)** Schematic of bottom data acquisition box

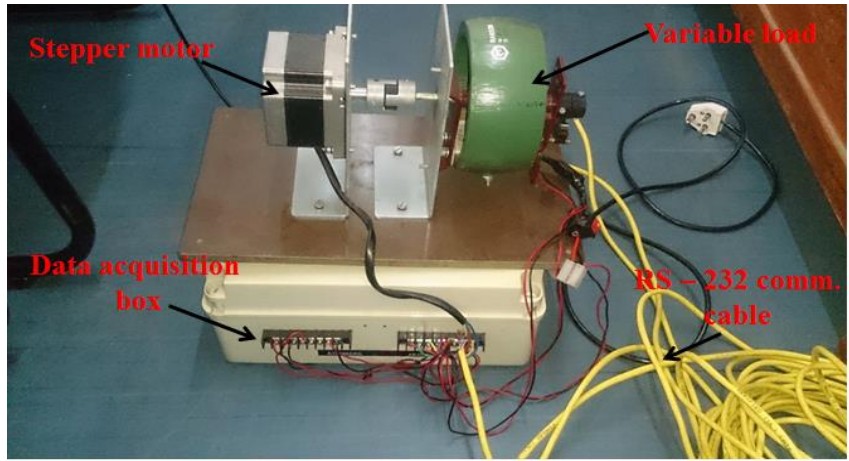

**Fig.4(b)** Bottom integrated data acquisition and control unit

Fig. 4(c) depicts the graphical user interface at the connected laptop to implement scanning window technique for maximum power point tracking of the array and visualize the sensors data.

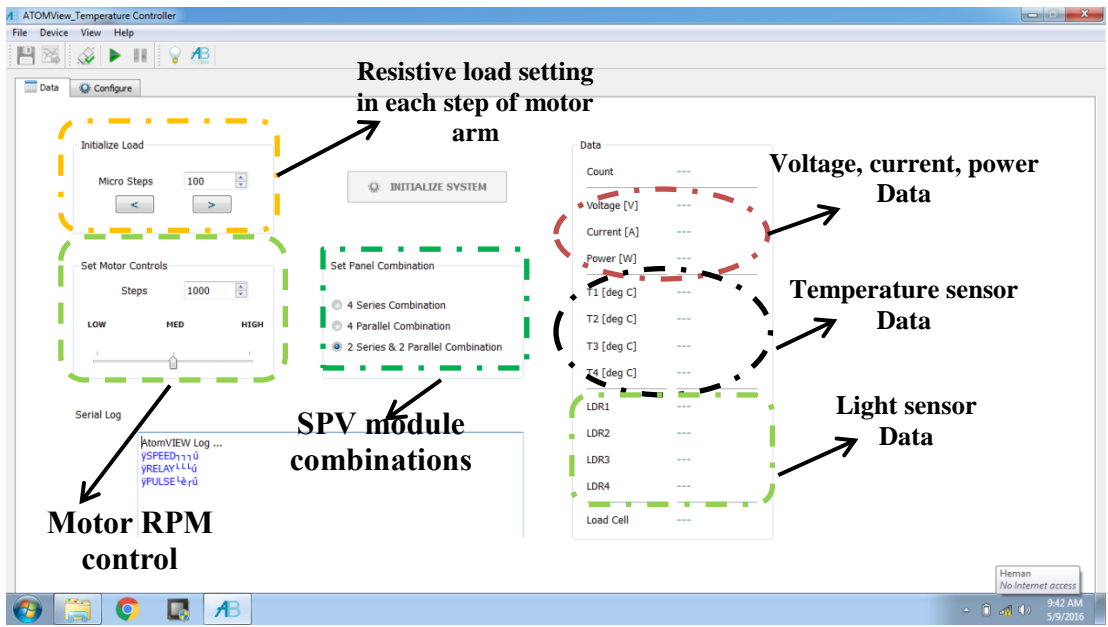

**Fig.4(c)** Graphical user interface of data acquisition and control unit

Solar output signal is given to the solar voltage input pins of the bottom board. Then solar output signal is given to the rheostat. Sensing register measures current and it is given to the solar current input. Microcontroller board also displays the speed and direction signal to the stepper motor driver. Stepper motor is used to vary the load resistance to obtain the maximum power point. Hence an option to control the motor RPM as well as setting the resistive load in each step of that motor is provided to track maximum power point of the SPV array effectively. One safety switch is provided to prevent short circuit. This entire process further helps in implementing scanning window technique (SWT) for maximum power generation.

In order to employ maximum power point tracking, the mounted solar PV modules have been configured in different combinations: series, parallel and mixed (series + parallel) combinations. Among these combinations it is found that series combination produces maximum power due to the voltage stability. Different combinations are being explained in Fig. 5.

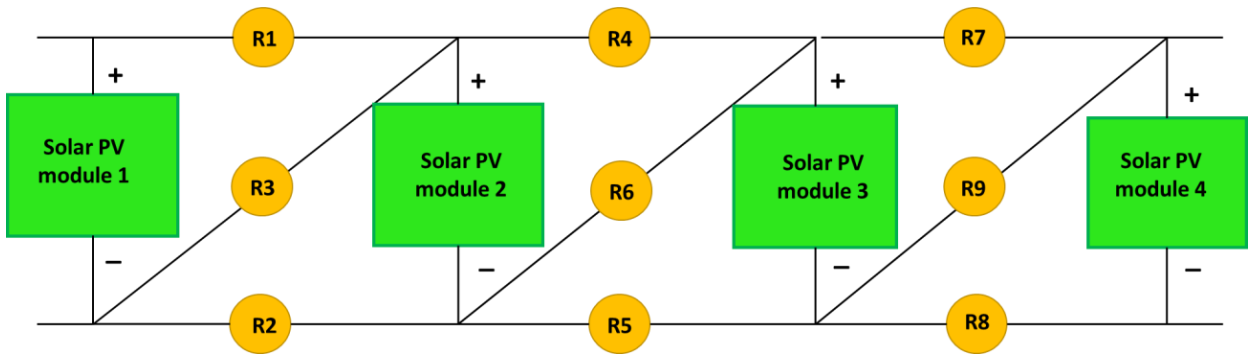

**Fig. 5** Configurations of SPV modules

Different combinations are being tried to get maximum power which are depicted below

**Parallel combination:** Only R1, R2, R4, R5, R7 and R8 are ON.

**Series combination:** Only R3, R6, R9 are ON.

**Series-Parallel combination:** Only R1, R2, R6, R7, R8 are ON.

But due to advantage of voltage and current stability in the PV system, series-parallel combination is being recommended which can enable the onboard instruments on the LTAP. In this research work primarily power generation characteristics on a LTAP using series-parallel combination is being discussed.

# RESULTS AND DISCUSSIONS

Power generation characteristic of flexible amorphous Silicon SPV modules on a contoured surface are well demonstrated by Sharma et al. [28. Sharma, 10] which is depicted figure 6. This exhibits almost 30% power loss in flexible photovoltaic (FPV) modules due to distributed non-uniform illumination on the contoured surfaces. Non-uniform illumination on a contoured surface is inevitable and cannot be ignored. Hence SPV modules integrated on a contoured surface are unable to generate maximum power as per the standards. Thus power generation of SPV modules on a contoured layout such as on lighter-than-air platform is directly dependent on its contoured shape which is demonstrated in this work.

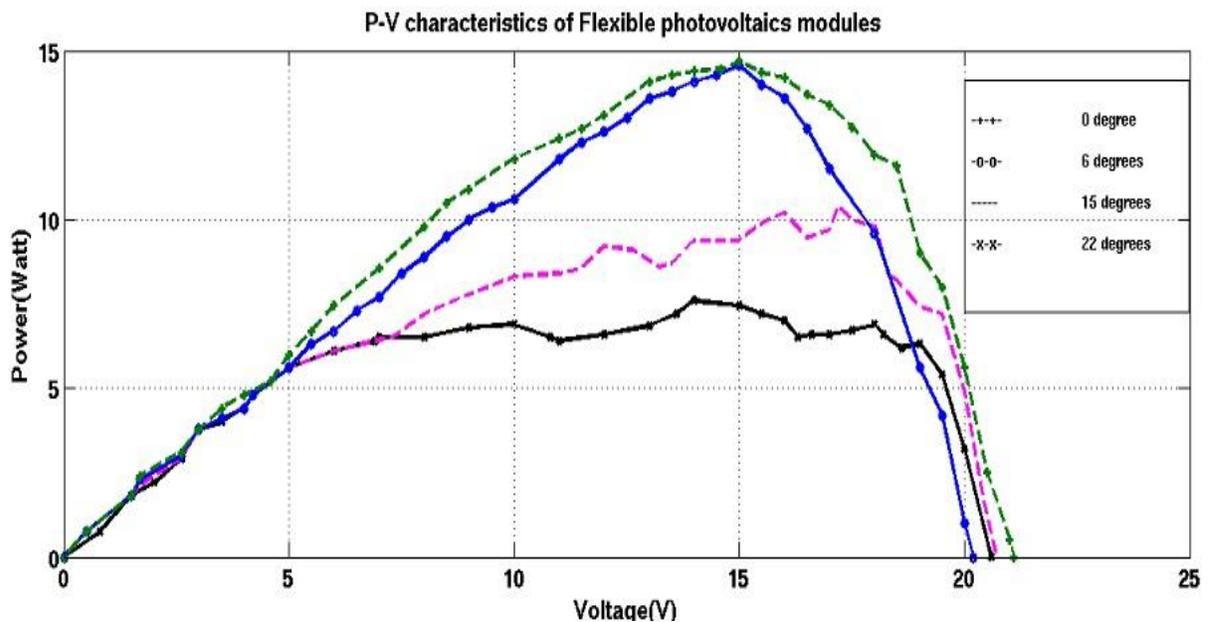

**Fig.6** P-V characteristics of contoured flexible PV modules [28. Sharma, 10]

Figure 6 shows that there is almost zero penalty in power generation up to $6^0$ contoured platforms. But beyond that, contouring has a considerable effect on generated power which leads to lower efficiency of the integrated SPV system. Mounted SPV modules at the $15^0$ contoured platforms show almost 28% of power loss compared to the mounted SPV modules between $0^0 – 6^0$ contour. This phenomenon leads to a detail study on the structure of the Lighter-than-Air platform structure. This structure is required to be aerodynamically stable and tethered to the ground. Iis filled with lighter-than-air or buoyant gas (such as hydrogen or helium) to generate an buoyant lift and to carry sophisticated instruments, sensor network for monitoring atmospheric environment, surveillance, reconnaissance, GIS etc. In order to facilitate power to those instruments, SPV modules are laid on the top of that platform. These platforms are not flat, rather round and contoured shaped in nature to attain a stable aerodynamics. The effective surface area for the SPV array layout needs to be derived for lossless power generation on the contoured LTAP. G.N.V.Rao shape has been considered for the constructing this lighter-than-air platform which is named after G.N.V Rao of Indian Institute of Science, who invented it. Its profile consists of ellipse, circle and parabolic shapes altogether across the length of the platform. The entire geometry of GNVR shape is parameterized in terms of maximum diameter, as shown in below fig.7.

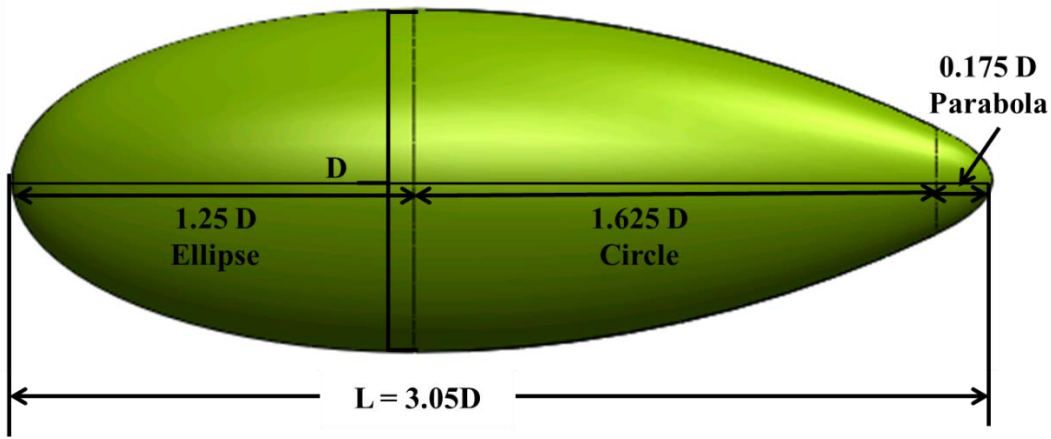

**Fig. 7** GNVR shape

The governing equations for each section are giving as follows [19. Ram, 10]:

$$\frac{x^2}{(1.25D)^2} + \frac{y^2}{0.5D^2} = 1 \qquad \text{------ (5)}$$

$$x^2 + (y - 3.5D)^2 = 16D^2 \qquad \text{------ (6)}$$

$$y^2 = 0.1373D(1.8D - x) \qquad \text{------ (7)}$$

This shape is a body of revolution which is a combination of elliptic, circular and parabolic profiles. Due to its axisymmetric nature, it has identical curved surface on both sides of the major axis. This contoured surface needs to be considered while planning the SPV array layout. Based on P-V characteristics of SPV modules on contoured surfaces (fig.6), it has been decided to keep the contour limit maximum within $15^0$ on both sides of the axis as shown in fig.8. Isometric cross-sectional view of the effective area for SPV array is represented in the fig. 8(b). This shows the effective surface area for SPV array layout on a LTAP. An analytical expression of effective surface area for each and every section and total effective surface area is given as follows.

$$A_{Ellipse} = \frac{\pi b}{6a} \left[ \frac{x\sqrt{a^2-x^2}}{2} + \frac{a^2}{2} \sin^{-1} \frac{x}{a} \right]_{-12.5D}^{0} \quad\text{------ (8)}$$

$$A_{Circle} = \frac{\pi}{6} \left[ \frac{x\sqrt{a^2-x^2}}{2} + \frac{a^2}{2} \sin^{-1} \frac{x}{a} - 3.5Dx \right]_{0}^{1.625D} \quad\text{------ (9)}$$

$$A_{Parabola} = \frac{\pi}{6} [0.155008064Dx]_{1.625D}^{1.8D} \quad\text{------ (10)}$$

So the total effective area is given by,

$$A_{Total} = A_{Ellipse} + A_{Circle} + A_{Parabola} \quad\text{------ (11)}$$

Corresponding volume of each sub-sections as well as volume of total LTAP is given from equations 12 to 15.

$$V_{Ellipse} = \pi b^2 \left[ x - \frac{x^3}{3a^2} \right]_{-1.25D}^{0} \quad\text{------ (12)}$$

$$V_{Circle} = \pi \left[ (\alpha^2 + \beta^2)x - \frac{x^3}{3} - \alpha x\sqrt{\beta^2 - x^2} - \alpha\beta^2 \sin^{-1} \frac{x}{\beta} \right]_{0}^{1.625D} \quad\text{------ (13)}$$

$$V_{Parabola} = \pi \left[ c\gamma x - c\frac{\gamma^2}{2} \right]_{1.625D}^{1.8D} \quad\text{------ (14)}$$

$$V_{Total} = V_{Ellipse} + V_{Circle} + V_{Parabola} \quad\text{------ (15)}$$

Where,

$a = 1.25D$

$b = D/2$

$\alpha = 3.5D$

$\beta = 4D$

$\gamma = 1.8D$

$c = 0.1373D$

The distribution of width on this effective surface area is not uniform across the entire geometry. It varies with respect to the length of the major axis as shown in fig.9.

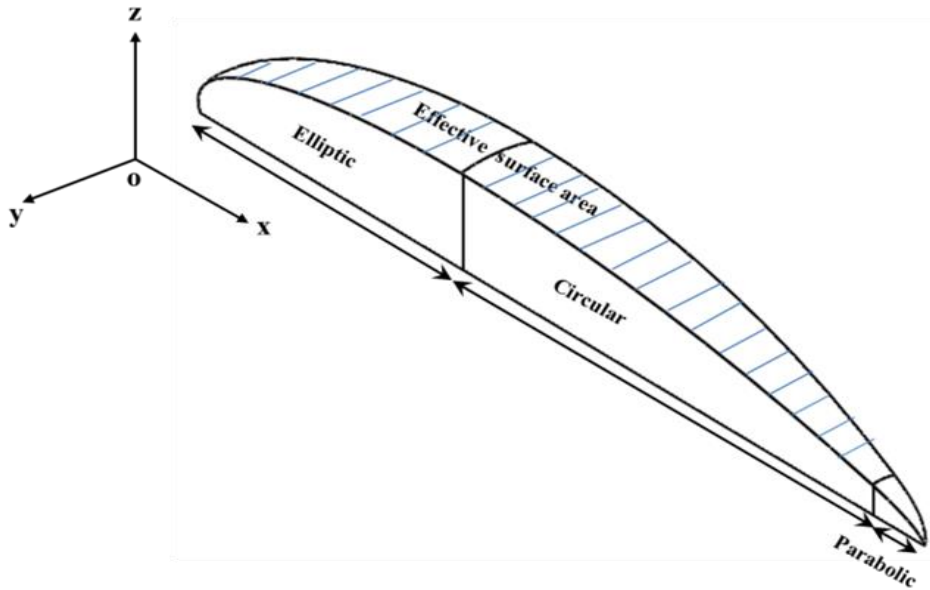

**8(a)** Schematic of effective surface area

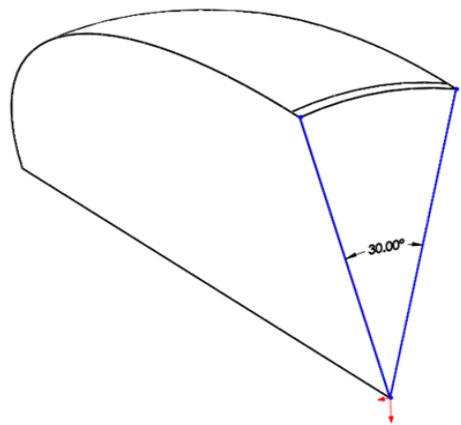

**Fig.8 (b)** Cross section of effective surface area

This restricts the number of SPV modules which can be mounted on the LTAP. Moreover it incurs in a loss in energy. Hence effective width plays a pivotal role while planning the SPV array layout. Based on figures of merit as referred in table 1, thin film crystalline silicon PV modules from High-Flex solar have been selected as power source. This SPV module has ~1 m width and ~1.94 m length. Thus the width of the effective surface area needs to be minimum 1 m to accommodate this SPV module.

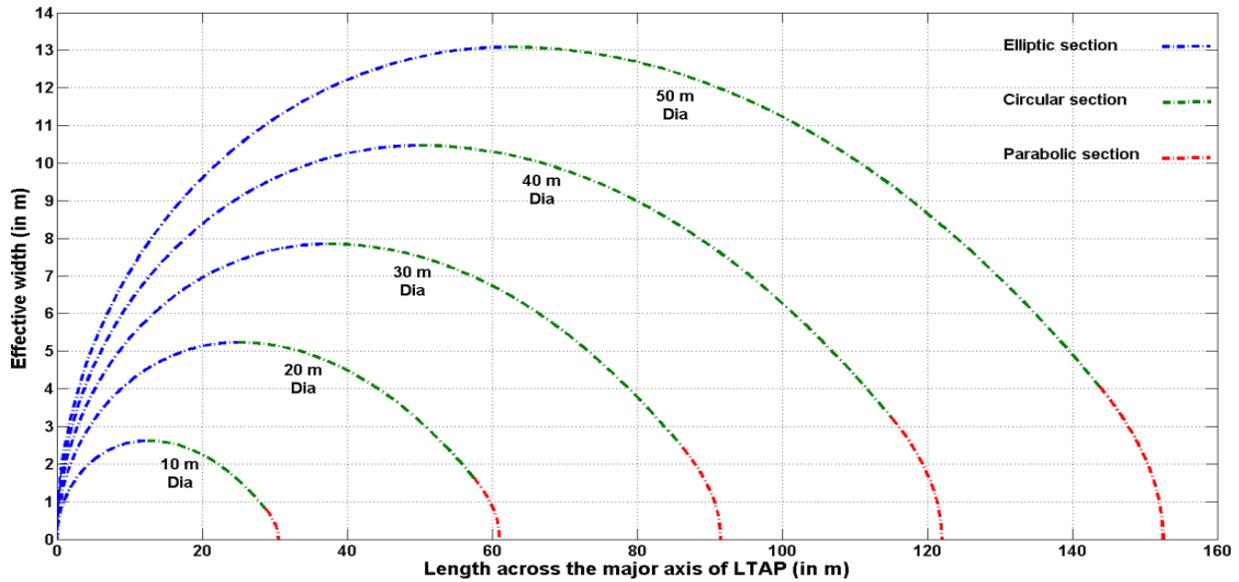

**Fig.9** Effective width across the major axis of LTAP

This width determines the entire architecture of SPV array layout using the dimension of that SPV module on the LTAP. Subsequently it determines the potential of the cumulative power generation by that LTAP.

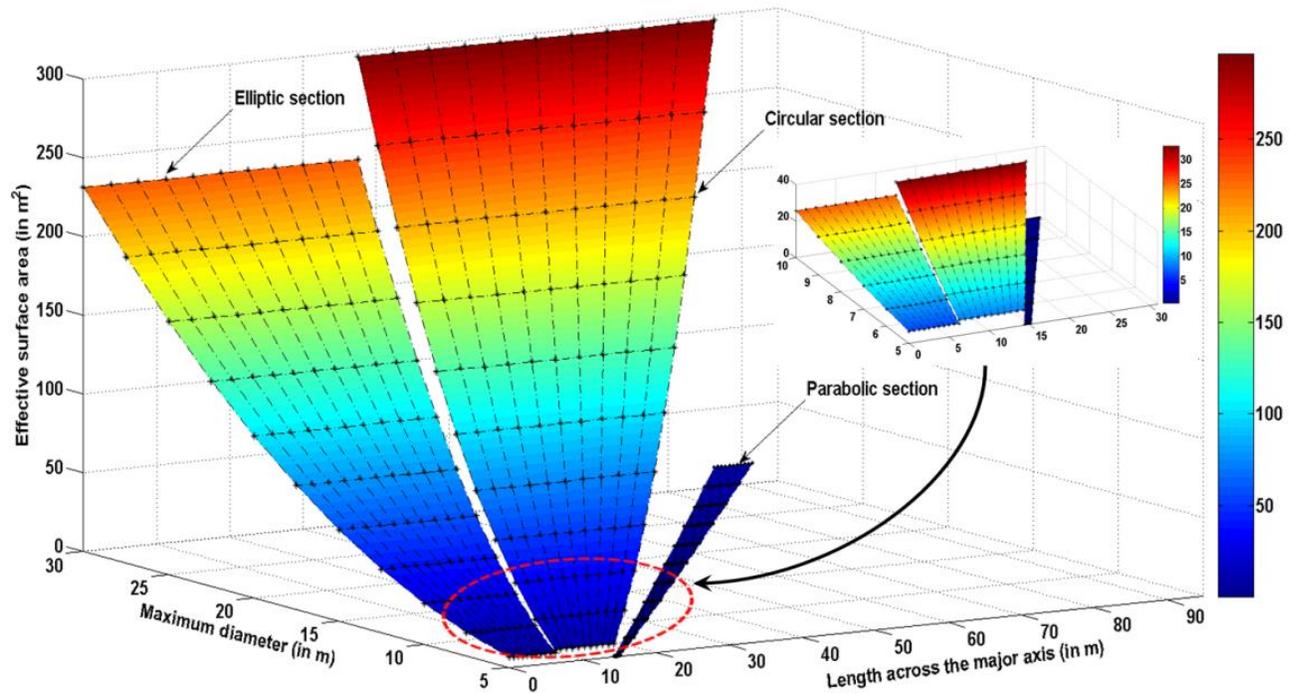
**Fig.10** Effective surface area across the major axis

Effective width is an important parameter that correlates the structural parameter with the potential power generation by that structure. Eventually the maximum diameter gets associated with the maximum power generation capacity through the effective surface area. This phenomenon constructs a function of a function which captures the concept of entire power generation in terms of maximum diameter of the LTAP. Fig. 10 shows the distribution of effective surface area for each and every section - elliptic, circular and parabolic. It is found that the circular profile contributes higher than the other sections. Thus most of the SPV modules can be integrated in this region. Fig. 11 provides 2-D interpretation of the effective surface area as well as quantifies the contribution by each section on that effective surface area. It shows that the circular region contributes 22% more area than the elliptic section and 95.7% more area than the parabolic region. The elliptic region contributes 94.33% more area than the parabolic region.

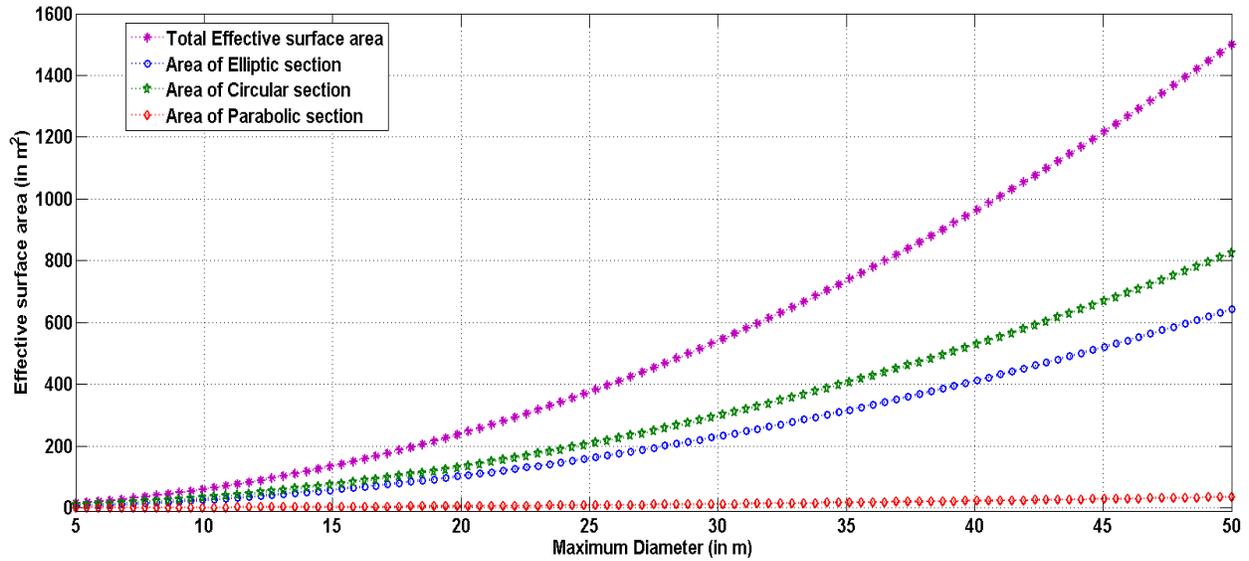

**Fig.11** Distribution of effective surface area across the major axis

Fig. 12 shows that 55% of effective surface area lies in the circular region, 43% in the elliptic region and 2% in the parabolic region. The entire effective surface area cannot be covered by the SPV modules due to its fixed length and width. This leads to an unutilized effective surface area while integrating that specific SPV module. This is further categorized in detail with respect to different diameter LTAPs in table 2. It demonstrates that unutilized effective surface area diminishes with increasing diameter of the LTAP.

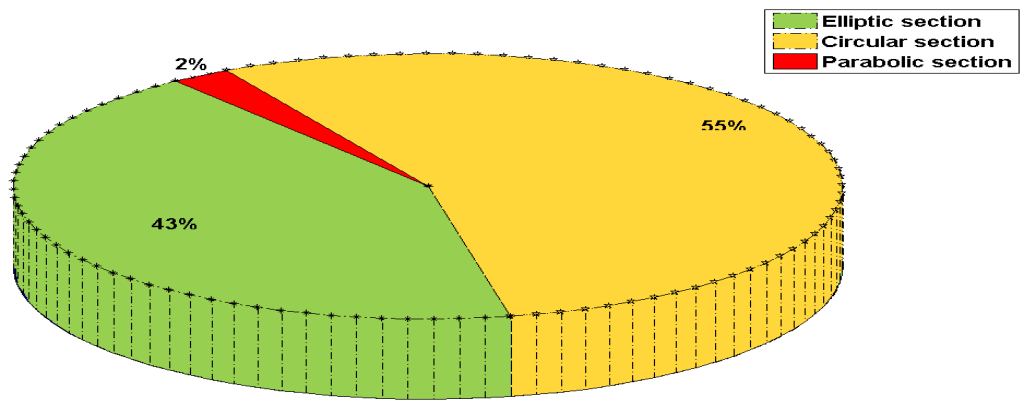

**Fig.12 Pie** chart of the effective surface area

These facts and figures are evident only for these particular SPV modules. Change in dimensional characteristics of the SPV module will lead to corresponding changes in the unutilized effective surface area. However, the trends are expected to remain same.

Table 2 : Utilization of effective surface area for LTAPs of different sizes

| Max. Diameter (in m) | No. of laid High-Flex solar PV modules | Max. Power generation by the laid SPV array (in kW$_p$) | Effective surface area (in m$^2$) | Unutilized surface area (in m$^2$) | Unutilized effective surface area |
|---|---|---|---|---|---|
| 5 | 4 | 2.52 | 15.003 | 7.003 | 46.68% |
| 10 | 19 | 5.985 | 60.02 | 22.02 | 36.69% |
| 15 | 51 | 16.065 | 135.03 | 33.03 | 24.47% |
| 20 | 93 | 29.295 | 240.05 | 54.05 | 22.5% |
| 25 | 147 | 46.305 | 375.082 | 81.082 | 21.62% |
| 30 | 223 | 70.245 | 540.1 | 94.1 | 17.43% |

The loss in maximum power generation capacity by LTAPs of different diameters due to unutilized effective surface area, is shown in fig.13. This gives a direct relationship between maximum power loss and maximum diameter of LTAP for an SPV module of this size.

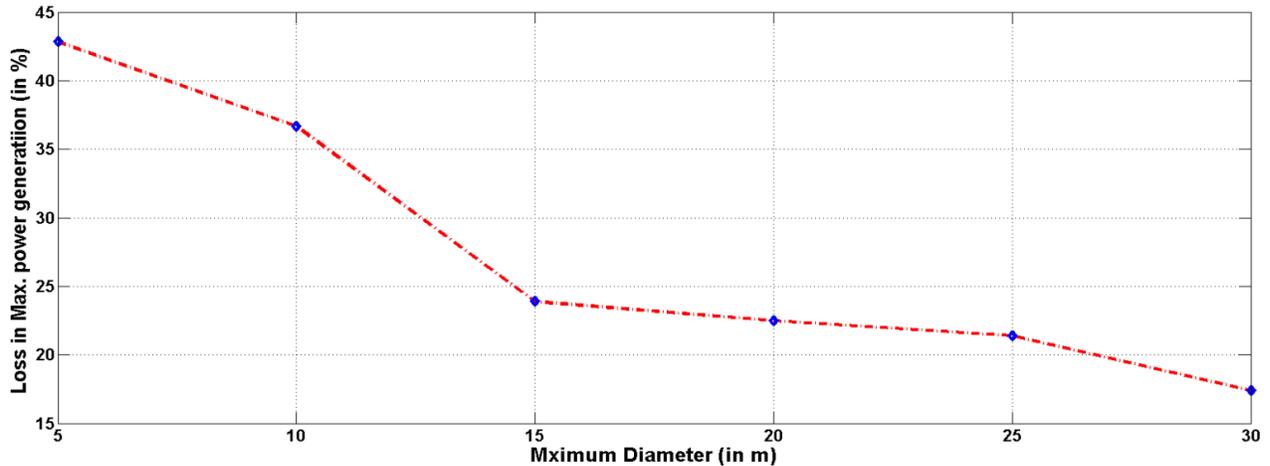

Fig.13 Losses in max power generation on a LTAP

It envisages that 5m diameter LTAP would experience 42.8% losses in maximum power generation capacity in opposition to 17.4% losses by the 30m diameter LTAP. The loss in maximum power generation capacity shows a decreasing trend with increasing maximum diameter. The corresponding volume of the LTAP also plays an important role in determining the payload ability. Payload comprises of envelope, ballonets, wings, tether, power generation unit (mounted SPV modules, power cables) and other sub-systems (instrumentation box, sophisticated camera, portable synthetic aperture radar etc.). LTAP volume determines the buoyancy which can support the required payload for maximum power generation. Hence LTAP volume, which is a function of maximum diameter, is also indirectly related to the power generation capability. Figure 14 depicts the contribution of each section of the LTAP to its total volume. The pattern is the same as that for the effective surface area. Circular section contributes the most followed by elliptic and parabolic.

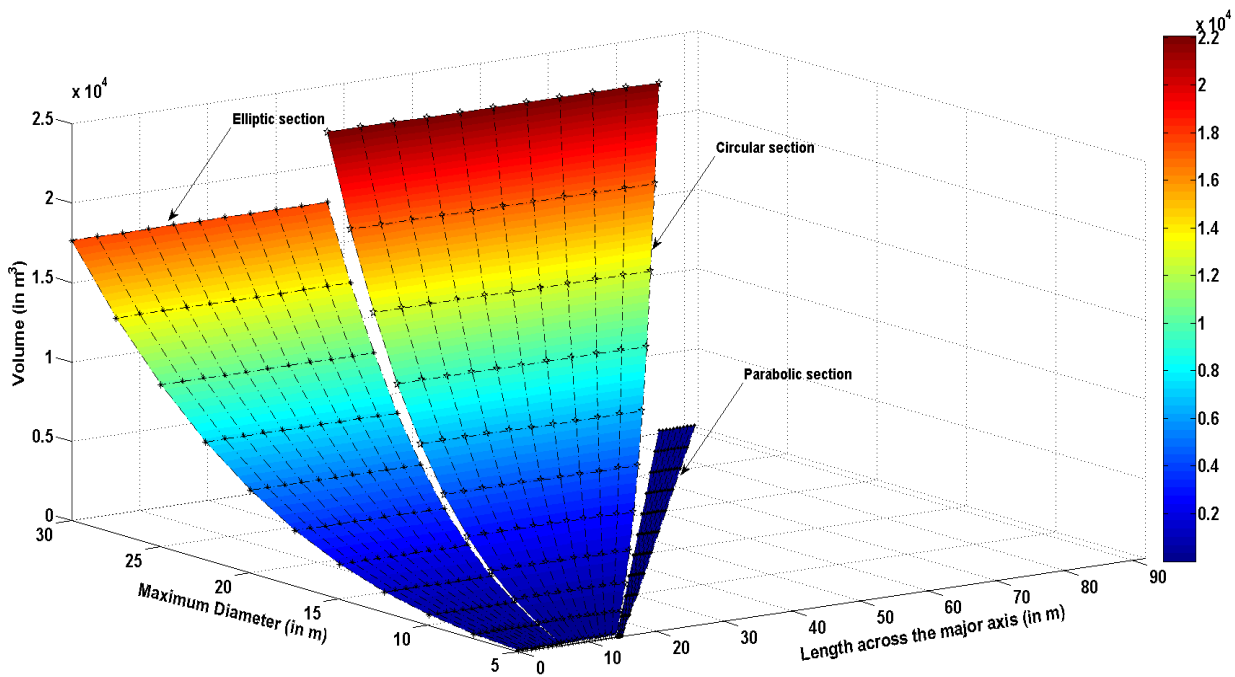

**Fig.14** Volume of the different sections of the LTAP

Circular section offers 20% more volume than elliptic section whereas parabolic section offers 98.9% and 99.2% less than the elliptic and circular sections respectively. Fig.15 gives another pictorial depiction of these volumetric contributions.

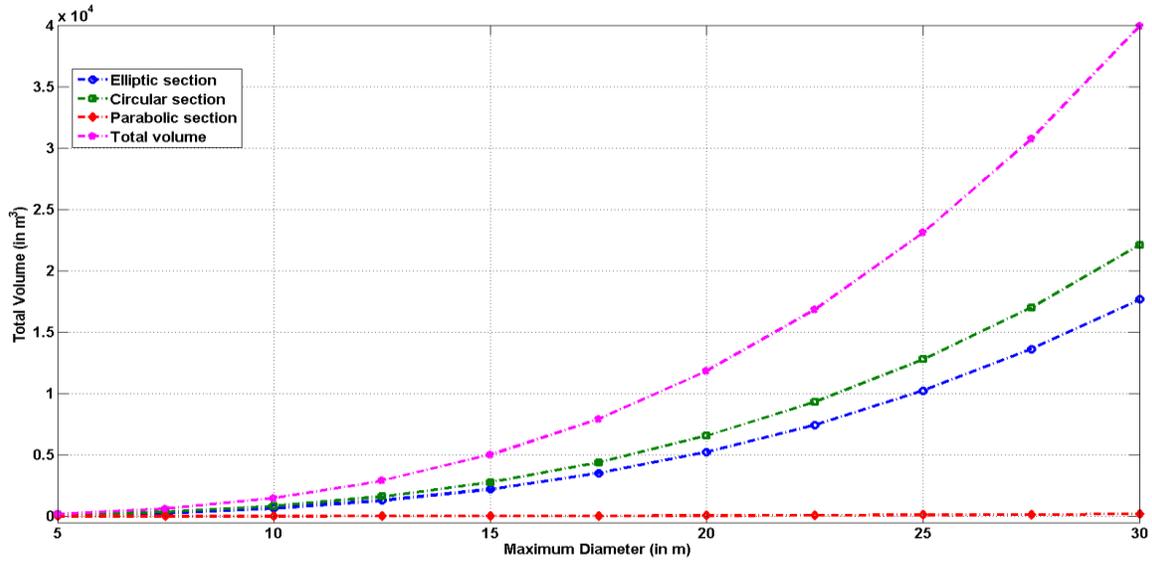

**Fig.15** Volume of the different sections of the LTAP

The pie chart in Fig. 16 provides the percentages contributed by each section to the total volume of the LTAP. This shows that contribution of the circular section, elliptic section and parabolic section are 55%, 44% and 1% respectively.

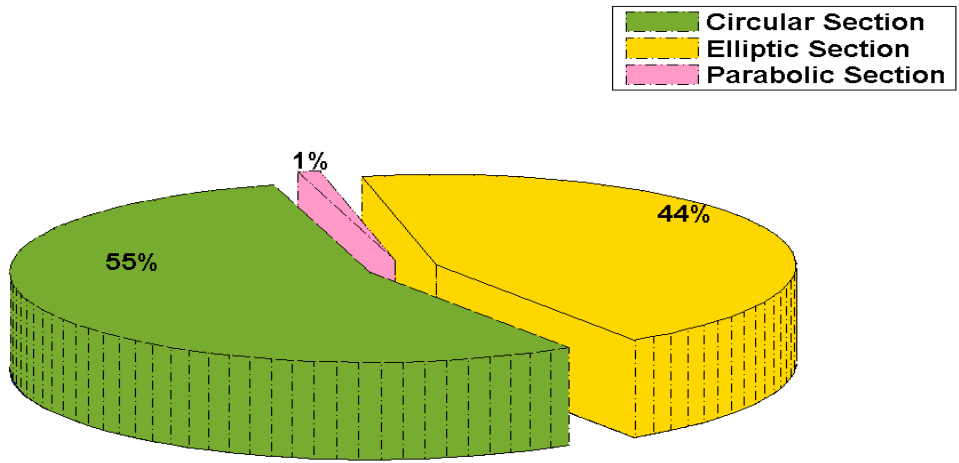

**Fig.16** Pie chart of the volume

Incident solar insolation on a contoured surface is non-uniform in nature. This phenomenon is depicted in figure 17. This result is obtained while positioning a SPV-LTAP on north-south direction at 12 noon at 0.5 km altitude from the mean sea level in Mumbai (latitude $19^0 07'$ N, longitude $72^0 51'$ E). Surface area with zero contours is receiving maximum solar insolation compared to rest of the surface. It is clearly visible that due to varying surface contour, non-uniform solar insolation is taking place. Henceforth SPV modules are integrated on the LTAP at the minimum contoured surface in order to receive maximum solar insolation. The threshold for integrating SPV modules is kept $15^0$ on the both sides of the zero contoured regions.

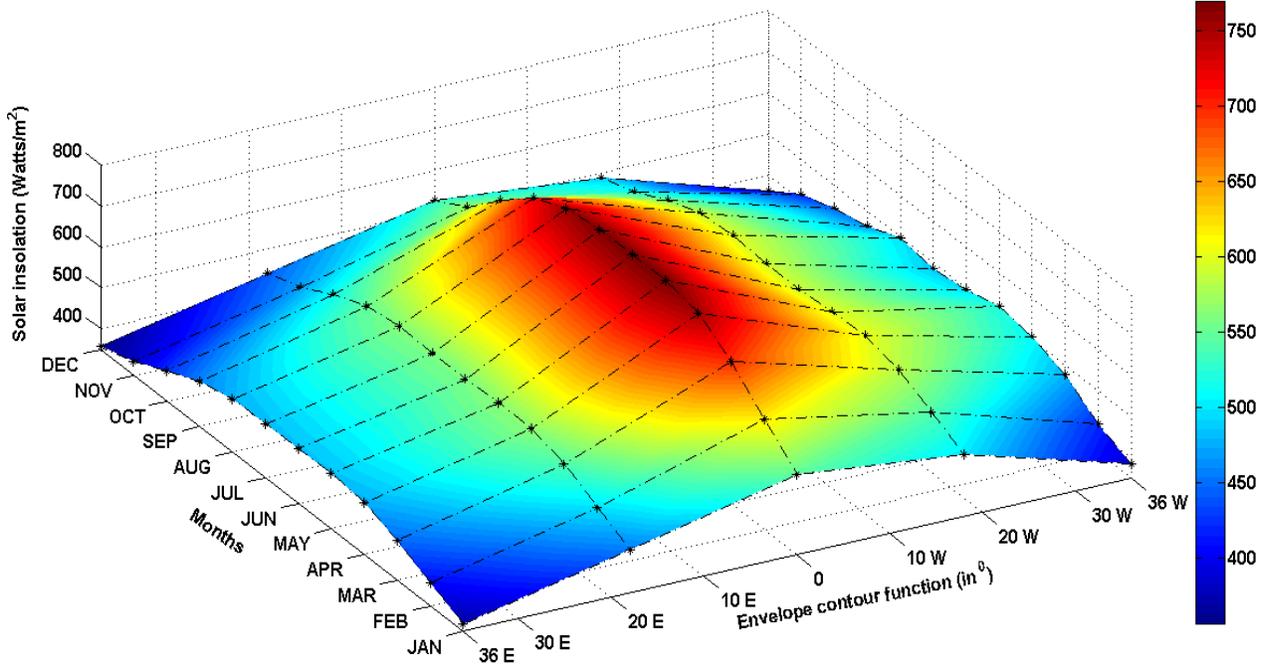

**Fig.17** Incident solar insolation on a contoured surface (Watts/m$^2$)

This characteristic of incident solar insolation on an LTAP, leads to a functional relationship with structural parameter i.e. surface contour function. Further, this surface contour function is dependent on the maximum diameter (D). This incident solar insolation flux on a tilted surface is governed by a parametric equation which is given as [37. Sukhatme, 08]

$$I_T = \sum_{i=0}^{N} I_b r_b + I_d r_d + (I_b + I_d) r_r \qquad \text{------ (16)}$$

Where  $I_T$ = Total solar insolation flux on the contoured surface

N = no of discrete contours

$I_b$ = Direct solar insolation flux

$I_d$ = Diffused solar insolation flux

$r_b$ = Beam radiation factor

$r_d$ = Diffused radiation factor

$r_r$ = Reflected radiation factor

Further these terms can be expressed as,

$$r_b = \frac{\omega_{st} \sin\delta \sin(\Phi-\beta) + \cos\delta \sin\omega_{st} \cos(\Phi-\beta)}{\omega_s \sin\Phi \sin\delta + \cos\Phi \cos\delta \sin\omega_s} \quad \text{------ (17)}$$

$$r_d = \frac{1+\cos\beta}{2} \quad \text{------ (18)}$$

$$r_r = \frac{\rho(1-\cos\beta)}{2} \quad \text{------ (19)}$$

$$\text{Declination } (\delta) = 23.45 \sin\left[\frac{360}{365} * (284 + n)\right] \quad \text{------ (20)}$$

Where n = day of the year

β = contour angle

Φ = latitude of the location

ρ = reflectivity of the surface

$\omega_s$ = sunrise/sunset hour angles for a horizontal surface

$\omega_{st}$ = sunrise/sunset hour angles for a tilted surface [37].

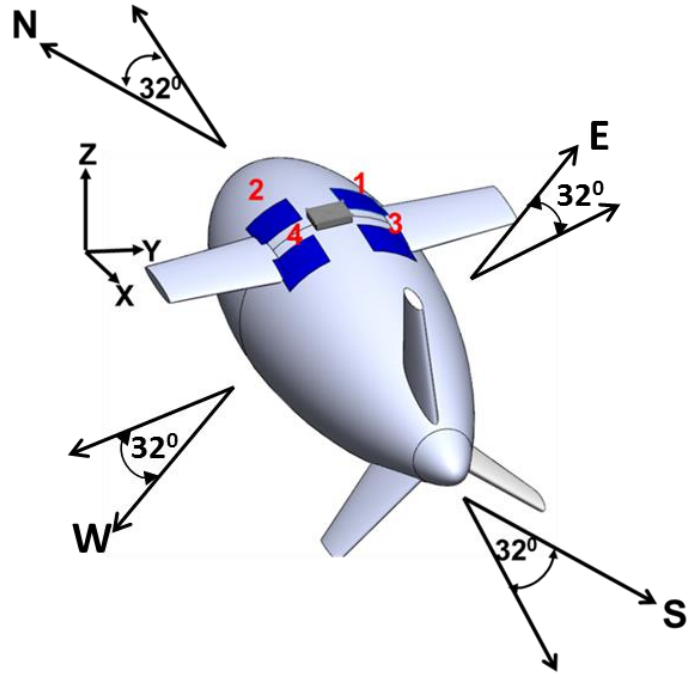

**Fig.18 (a)**

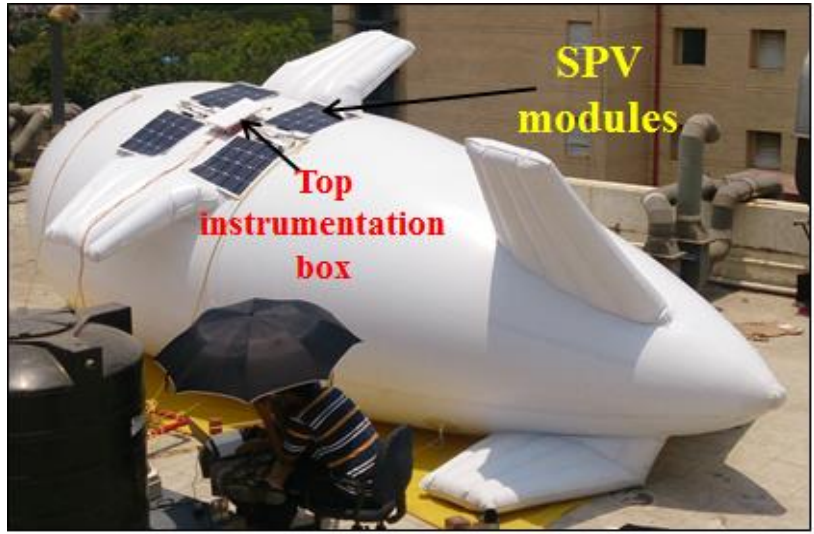

**Fig.18 (b)**

**Fig.18 (a)** Positioning of the experimental model, **(b)** Experimental model

Figures 18(a) and 18(b) explain the positioning of the experimental setup at $32^0$ azimuthal angle with the N-S direction. Due to this azimuth angle, incident solar irradiance on the LTAP reduces which is seen in figure 19.

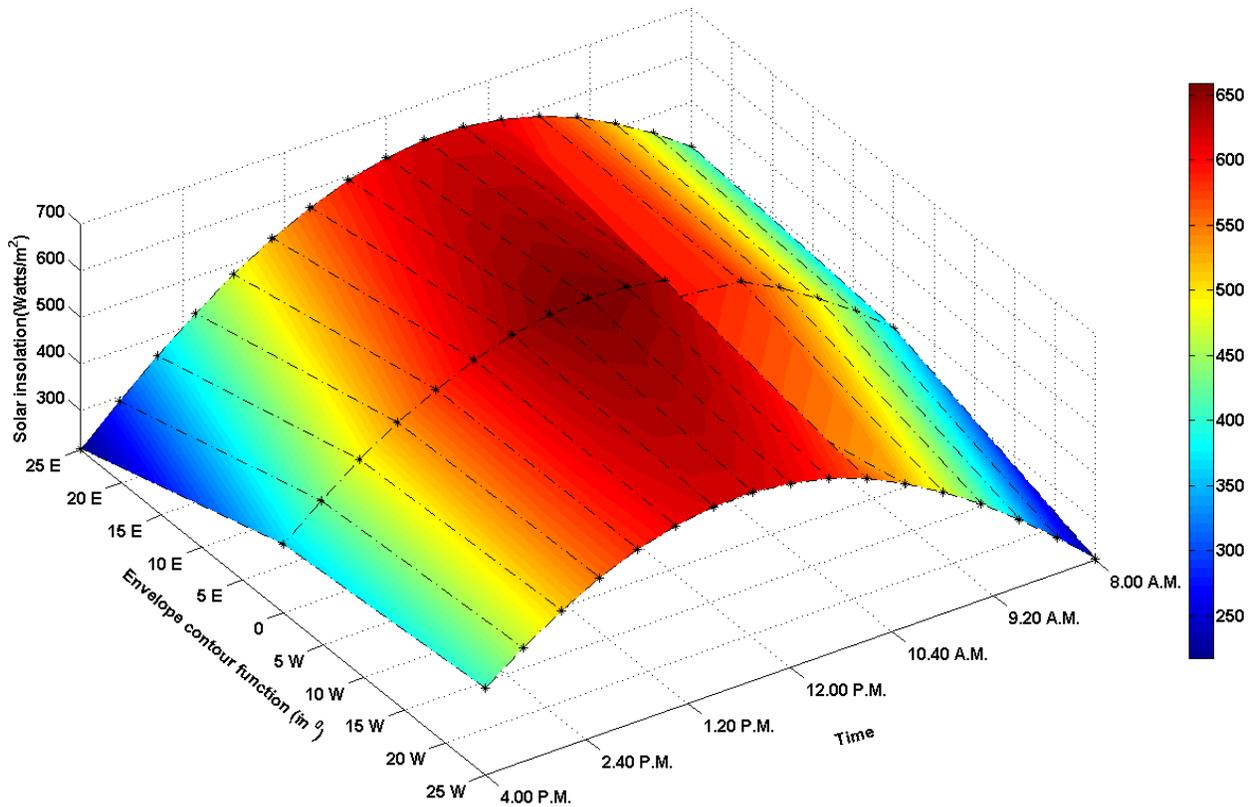

**Fig.19** Incident non-uniform solar insolation on the experimental model (Watts/m$^2$)

Azimuth angle produces a discontinuity in the pattern of incident solar irradiance on the mounted SPV array. This happens while shifting the positioning of the sun from west to east during the course of day. It is evident that SPV modules mounted on zero contoured area receives maximum incident insolation over the entire day. This establishes the fact of inevitable dynamic variable insolation on the mounted SPV array.

The HF-40 SPV modules consist of 16 solar cells in a 4x4 array. Each string comprising of 4 solar cells can be considered a sub-module. Thus every HF-40 SPV module is an integration of 4 sub modules. When these HF-40 SPV modules are integrated on the experimental LTAP, it consumes $13.5^0$ to $38.5^0$ of the envelope contour function. Each and every string of the SPV module accounts for $6^0$ of the envelope contour function. This is depicted figure 19. It is evident that each sub-module would receive different insolation values. However, to simplify the calculations, incident solar insolation at the $25^0$ envelope contour function assumed to be the distributed solar insolation throughout the array. In reality, the differences in insolation would results hot spots or localized heating in each and every sub-module within a module. This mismatch regarding the incident solar insolation on each string of the SPV module can be mitigated by integrating bypass diodes across the each sub module of the array. But only one or two bypass diodes are integrated with each commercially available SPV modules to reduce the cost. HF-40 SPV modules have a single bypass diode for the entire module. The I-V and P-V characteristics recorded by the experimental setup are given in figures 20 and 21. This shows that this array produces maximum power at 12 noon and minimum power at 10:00 AM. Due to change in dynamic positioning of the sun, incident insolation on the array changes with respect to time. This accounts for different power generation at different times. The incident insolation results reported earlier shows that the SPV array would receive maximum insolation between 11 am and 1.30 pm. It is found that current generated by SPV modules are directly proportional to the incident insolation on the SPV modules. The maximum current value of 3.77 A is available at the time of maximum insolation.

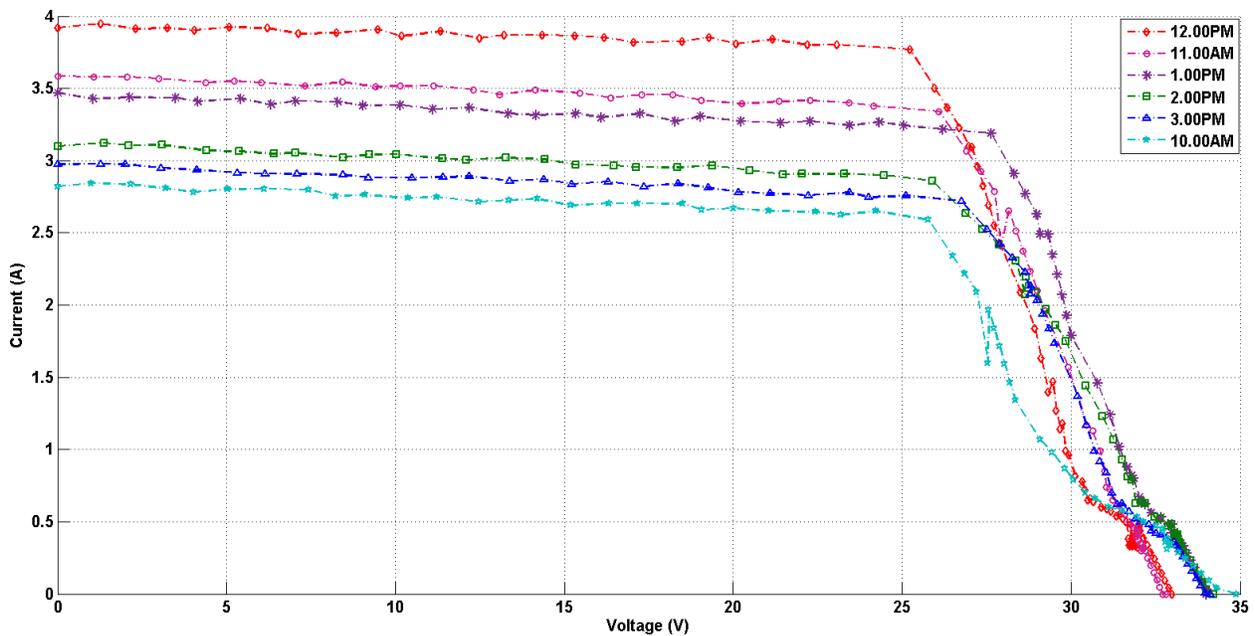

**Fig.20** I-V characteristics of the SPV integrated LTAP

On the other hand, the P-V characteristic result exhibits presence of some prominent ripples and multiple peaks in the power values. This is a result of variable current due to incident non uniform insolation at different contours of the LTAP. But ripples may also be due to relative displacement of the LTAP under wind induced vibration. Figure 21 depicts that the maximum and minimum power generated by this experimental model would be 95.12 Watts at 12.00 noon and 66.77 Watts at 10.00 AM respectively. Though this integrated power system generates maximum power at 12.00 noon, efficiency of the system is found to be highest at 1 pm.

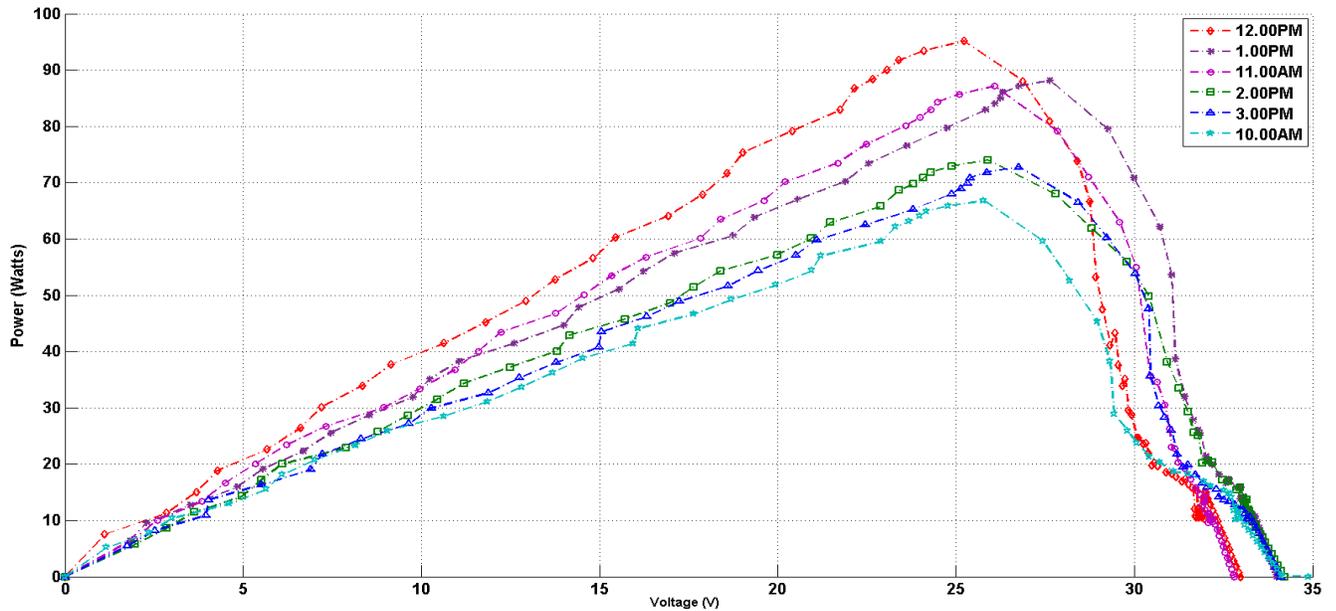

**Fig.21** P-V characteristics of the SPV integrated LTAP

Amongst the different maximum power point tracking (MPPT) techniques, scanning window technique (SWT) is being implemented in this work. The scanning window for this SPV array is derived from the fact that the MPP occurs at different percentages of open circuit voltage for this specific module technology, i.e., thin-film crystalline Si and is also sensitive to non-uniform insolation due to contoured structure in this case. [define Vfactor] In this experimental SPV array, $V_{factor}$ values can be easily calculated to formulate the scanning window power optimization. $V_{factor}$ values are found to very between **0.6 $V_{oc}$** to **0.82 $V_{oc}$** and this can be used to eliminate the local maximum power points at different times on the day. This technique keeps the operating points close to the global MPP (with no other peaks expected to be encountered). The width of the window changes with the different contoured structure due to change in incident insolation pattern. In figure 22, if the width of the scanning window is 2 V then it ranges from 4.7% of $V_{oc}$ to 6.3% of $V_{oc}$ at different times of the day. This SWT is a fast and efficient technique for obtaining the maximum power point for this contoured SPV array with minimal losses.

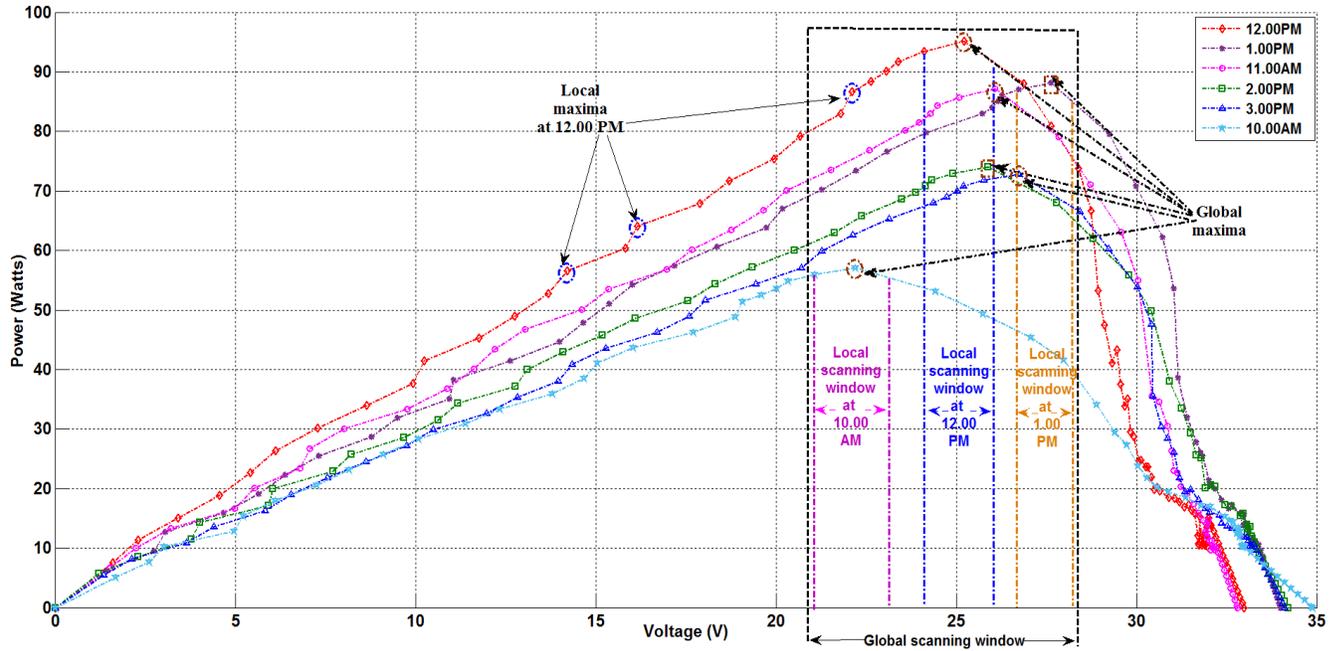

**Fig.22** P-V characteristics demonstrating the SWT for MPPT

Here efficiency of this power system is denoted in terms of the fill factor. This is defined as the ratio of maximum power generation to the product of $V_{oc}$ and $I_{sc}$. It is found that at 1.00 PM, this system attains a maximum fill factor of 0.747, in spite of generating maximum current ($I_m$) lesser than that at 11 AM and 12 noon (figure 23). This can be attributed to the highest maximum voltage ($V_m$) generation at that time. Thus it can be concluded that fill factor is directly proportional to the maximum voltage of the array. So a steady maximum voltage is expected to attain highest efficiency even under the presence of non-uniform insolation. Bypass diode integrated modules are unable to control the effect of non-uniform insolation as it considers the lowest incident insolation across the different sub modules. It does not consider the different insolation values across the module at the different contour regions.

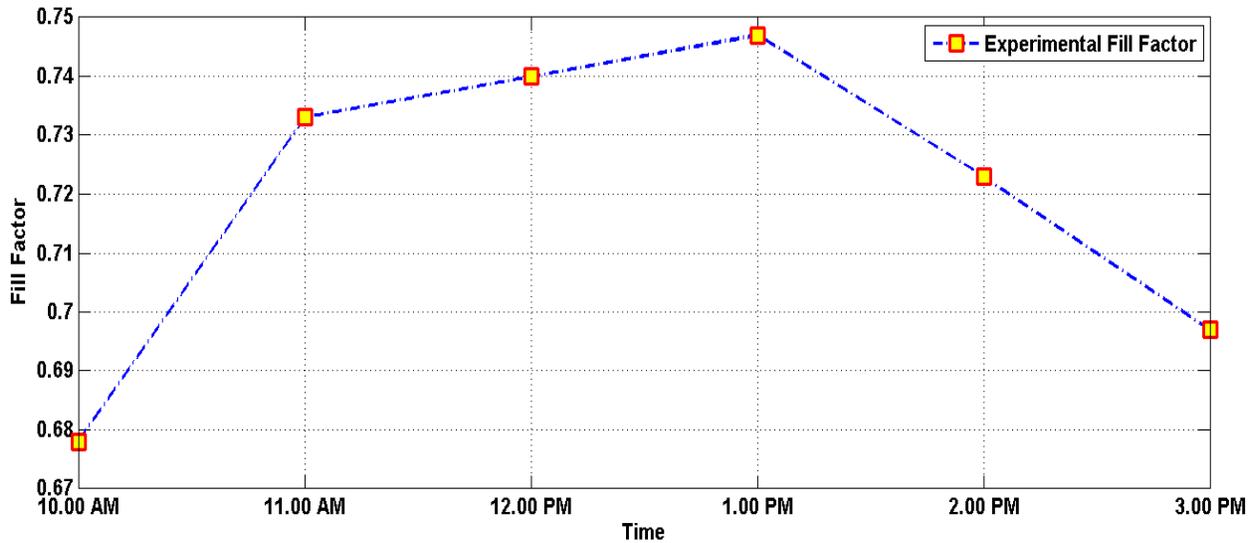

**Fig.23** Efficiency of the SPV integrated LTAP

Distributed maximum power point tracking (D-MPPT) algorithm is implemented in order to capture the effect of the variable insolation across the each and every sub-module. This can be attained by integrating the DC-DC converter across the each sub-module within a module as shown in figure 24. Earlier it has been mentioned that bypass diodes can be used to check these multiple peaks due to non-uniform insolation. But this simultaneously reduces the operational maximum voltage of the entire module and subsequently that of the array.

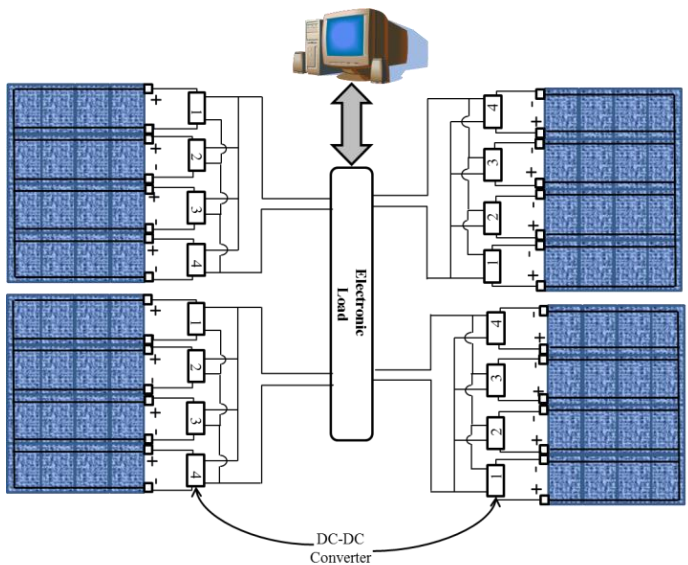
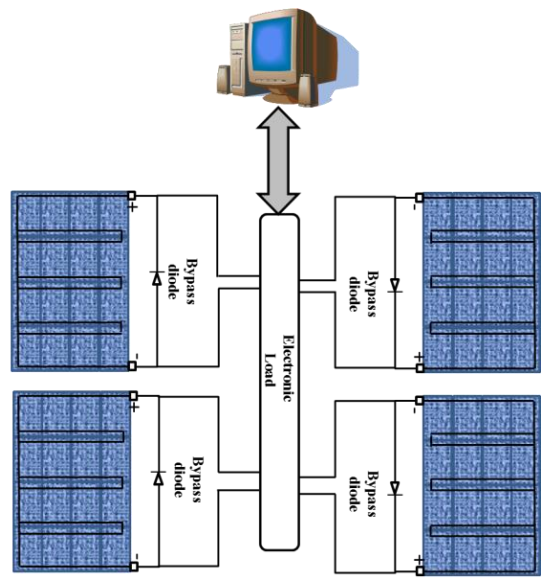

**Fig.24 (a)** Implementation of D-MPPT         **Fig.24 (b)** Conventional MPPT scheme

If bypass diodes are exploited, the four sub-modules are connected in series and the power is evaluated as four times the minimum substring power. In contrast, if DMPPT is used, the total power will be the sum of the power extracted from each sub-module. As the module power at MPP is approximately proportional to the incident insolation, it is possible to estimate the theoretical efficiency between the MPP power with bypass diodes and with DMPPT.

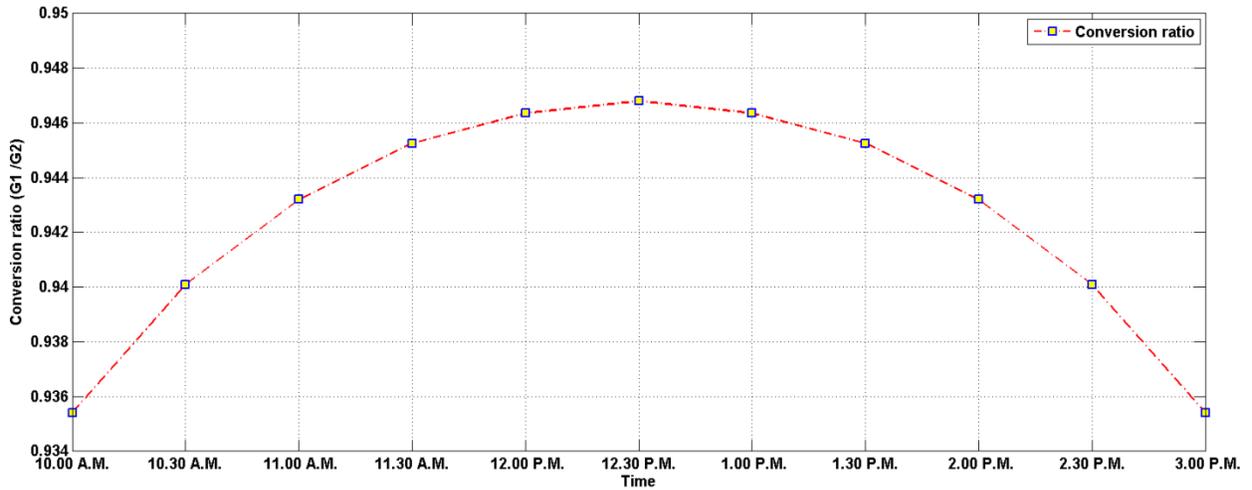

**Fig.25** Efficiency of the DMPPT algorithm

For example let us consider at a certain time an east facing module receives insolation values $G_{E1}$, $G_{E2}$, $G_{E3}$ and $G_{E4}$ Watts/m² ($G_{E1}>G_{E2}>G_{E3}>G_{E4}$) respectively for the four sub modules. Similarly a west facing module receives $G_{W1}$, $G_{W2}$, $G_{W3}$ and $G_{W4}$ Watts/m² ($G_{W1}<G_{W2}<G_{W3}<G_{W4}$) at four sub-modules. Therefore the power of the entire array using bypass diode based conventional MPPT scheme would be proportional to the

$$G_1 = 2(4G_{E4} + 4G_{W1}) \quad \text{------ (21)}$$

While the power under the DMPPT scheme would be proportional to the

$$G_2 = 2[(G_{E1} + G_{E2} + G_{E3} + G_{E4}) + (G_{W1} + G_{W2} + G_{W3} + G_{W4})] \quad \text{------ (22)}$$

The ratio between $G_1$ and $G_2$ denotes the power extraction using two MPPT schemes. It is found that sub-module integrated converters under the DMPPT scheme is more effective for drawing optimized power from the SPV integrated Lighter-than-air platform. These values may differ due to dynamic incident insolation over a period of time and the conversion losses which need to be explored further. On the contrary implementation of this DMPPT scheme will increase the initial cost of the system.

This entire work was aimed at developing a SPV integrated LTAP for optimized mobile power generation for a given volume. If the LTAP volume is known, then the corresponding payload of the integrated power system can be estimated. This will lead to the corresponding effective surface area and effective width on the LTAP for mounting the SPV array. This will provide the potential power generation ability of that LTAP which can be further optimized by implementing the scanning window (SW) MPPT technique. This complete process sequence is depicted in the flow chart given in figure 26.

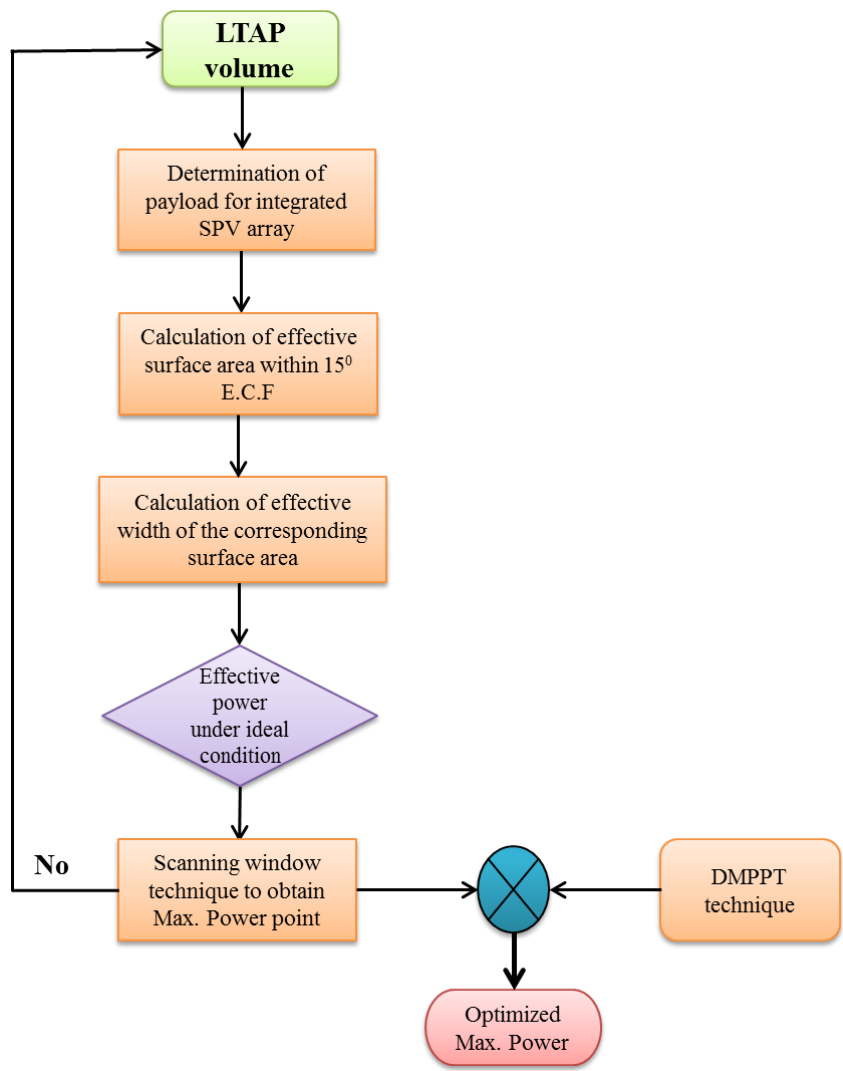

**Fig.26** Flow chart for development of a SPV integrated LTAP for maximum power generation

## CONCLUSION

This work describes the developmental algorithm for optimized power generation on a SPV integrated LTAP. It first describes the methodology to calculate the effective surface area for the SPV layout on the LTAP and its corresponding volume. Volume of the LTAP determines the payload ability and its sustainability at an altitude for a certain period of time. On a whole it helps to determine the energy generation capability for this aerial mobile power system. Non-uniform insolation is the contoured surface of the LTAP is quantified. The effect of non-uniform insolation is studied in detail through both experimental and theoretical results. Based on these non-uniform insolation characteristics, an SPV layout is planned and LTAP is oriented on a particular relative position with respect to sun to receive maximum insolation throughout the day. An experimental model of 24 $m^3$ LTAP was used as a prototype and HF-40 thin film crystalline Si modules are integrated on that platform to characterize the solar power generation on that LTAP. A significant loss is observed in power generation due to non-uniform insolation on the SPV array. Distributed MPPT method is proposed to optimize the power generation on the LTAP. By integrating sub-module integrated converter within the modules, power optimization can be achieved. This can cause an increase in power generation of about 10% compared to a conventional SPV array on an LTAP operated at its MPP.